\documentclass[12pt]{article}
\usepackage{epsfig,pstricks}

\newcommand{\nc}{\newcommand}
\nc{\beq}{\begin{equation}}
\nc{\eeq}{\end{equation}}
\nc{\beqa}{\begin{eqnarray}}
\nc{\eeqa}{\end{eqnarray}}

\newwrite\ffile\global\newcount\figno \global\figno=1

\def\writedef#1{}

\input epsf
\def\figin{\epsfcheck\figin}\def\figins{\epsfcheck\figins}
\def\epsfcheck{\ifx\epsfbox\UnDeFiNeD
\message{(NO epsf.tex, FIGURES WILL BE IGNORED)}
\gdef\figin##1{\vskip2in}\gdef\figins##1{\hskip.5in}
\else\message{(FIGURES WILL BE INCLUDED)}%
\gdef\figin##1{##1}\gdef\figins##1{##1}\fi}

\def\figinsert{}
\def\ifig#1#2#3{\xdef#1{fig.~\the\figno}
\writedef{#1\leftbracket fig.\noexpand~\the\figno}%
\figinsert\figin{\centerline{#3}}\medskip\centerline{\vbox{\baselineskip12pt
\advance\hsize by -1truein\center\footnotesize{  Fig.~\the\figno.} #2}}
\bigskip\endinsert\global\advance\figno by1}
\def\endinsert{}

\begin{document}

$\left. \right.$ \vskip 1.5 in
\begin{center}
{\large{\bf  On the QCD Ground State at High Density} }
\end{center}

\begin{center}
Nick Evans$^1$,
James Hormuzdiar$^2$,
Stephen D.H. Hsu$^3$,
Myck Schwetz$^4$

{\small\em {}$^1$Department of Physics,
University of Southampton, Southampton, S017 1BJ, UK.} \\

{\small\em {}$^2$Department of Physics,
McGill University, 3600 University,
 Montreal, Quebec, H3A 2T8, USA.} \\

{\small\em {}$^3$Department of Physics,
University of Oregon, Eugene OR 97403, USA} \\

{\small\em {}$^4$Department of Physics,
Boston University, Boston, MA 02215, USA.}  \\

\end{center}

\author{}

\begin{center}October, 1999
\end{center}


\begin{picture}(0,0)(0,0)
\put(350,300){BUHEP-99-25}
\put(350,255){SHEP-99-16}
\put(350,270){OITS-681}
\put(350,285){McGill-99/31}
\end{picture}

\begin{abstract}
We investigate the possible ground states of QCD at asymptotic densities, 
where the theory is expected to exhibit color superconductivity. 
We characterize the color-flavor structure of possible diquark 
condensates, and find those that are energy extrema by solving the 
weak-coupling Dyson-Schwinger equations, including Landau damping 
and the Meissner effect. We show that, as previously anticipated, 
in the two flavor case the vacuum breaks SU(3) color to SU(2) and 
in the three flavor case the vacua with color-flavor locking (CFL) 
have the lowest energy. We identify a number of relatively flat 
directions in the potential along which the pattern of gauge symmetry 
breaking changes and parity is violated. We discuss possible 
phenomenological consequences of our results.
\end{abstract}

\newpage

\section{Introduction}

QCD at high density and low temperature is a color superconductor   
\cite{colorsuper}-\cite{HS} characterized by the formation of a diquark 
condensate
in the attractive $\bar{3}$ color channel. The condensation is analogous to 
Cooper pairing in ordinary superfluids or superconductors, and occurs even via
arbitrarily weak attractive interactions due to the presence of a Fermi surface
\cite{EHS,SW,FLRG}. Recently it was discovered \cite{Son,PR,HongSD,SWgap,HS}
that long range magnetic fluctuations
enhance the condensation, leading to a gap which behaves as 
\beq
\label{LP}
\Delta \sim \mu g^{-5} \exp \Bigl( - {3 \pi^2 \over \sqrt{2} g} \Bigr) ~~~
\eeq
in the weak coupling (small g, or large $\mu$) limit. 
In this limit (\ref{LP}) is
the gauge invariant, leading order result of a systematic 
expansion in powers of g.
The properties of the condensate are easy to determine in 
the case of two quark flavors.
Because the condensate occurs between pairs of either 
left (LL) or right (RR) handed
quarks in the J=L=S=0 channel \cite{HS}, and the $\bar{3}$ color
channel is antisymmetric, the quarks must pair in the 
isospin singlet (ud - du) flavor
channel. However, even in this case there is a subtlety, as
the relative color orientations of the LL and RR condensates are 
not determined by the usual leading order analysis. A misalignment
of these condensates violates parity,
and further breaks the gauge group beyond 
$\rm SU(3)_c \rightarrow SU(2)_c$.
As we will discuss below, an analysis of the Meissner effect is 
necessary to determine the relative orientation.
There are thus a number of unstable configurations of only slightly
higher energy with
different color-flavor orientations (and hence different symmetry
breaking patterns), leading to the possibility of
disorienting the diquark condensate. We include a discussion of possible
phenomenological signals associated with these phenomena.

The generalization to three flavors is far from straightforward. 
Again, one can show that
the condensate must occur in the J=L=S=0 and color $\bar{3}$ channel. 
The Pauli principle
then requires that the flavor structure again be antisymmetric 
$\sim ~(q_i q_j - q_j q_i)$, for quarks of flavor $i,j$. Thus, 
one can have combinations
of condensates which are in the $\bar{3}$ of both color 
and flavor $\rm SU(3)_L$
or $\rm SU(3)_R$. Due to the chirality preserving nature of 
perturbative gluon exchange, 
there is no mixing of LL and RR condensates, which form independently. 
One can immediately
see that there are a number of possibilities. For example, the 
condensates for the three flavors and both chiralities might all align
in color space, leading to an $\rm SU(3)_c \rightarrow SU(2)_c$ 
breaking pattern. A more
complicated condensate has been proposed \cite{ARW2,SW2} called 
Color Flavor Locking (CFL),
in which the $\bar{3}$ color orientations are ``locked'' to 
the $\bar{3}$ flavor orientation.
 
In this paper we determine the nature of the energy 
surface governing 
the various color-flavor orientations of the condensate. Let 
us begin by characterizing
the color-flavor configuration space of condensates. We 
consider the ans\"atz
\beq
\label{ansatz}
\Delta^{ab}_{ij}~{}^{L,R} ~=~ A^c_k ~{}^{L,R}~ 
\epsilon^{abc} \epsilon_{ijk} ~~~,
\eeq
where a,b are color and i,j flavor indices. L and R denote 
pairing between pairs of
left and right handed quarks, respectively. Under color and flavor A
transforms as
\beq
\label{Atrans}
A^{L} \rightarrow   U_c A^L V^L~~~,
\eeq  
where $U_c$ is an element of $\rm SU(3)_c$ and $V^{L}$ of 
$\rm SU(3)_{L}$. A similar equation holds for $A^R$. 
It is always possible to diagonalize $A^L$ by appropriate 
choice of $U_c$ and $V^L$:
\beq
\label{ansatz1}
A^L  ~=~ 
\left(\begin{array}{ccc}
a & 0 & 0 \\
0 & b & 0 \\
0 & 0 & c 
\end{array} \right)~~~.
\eeq
Generically, there does not exist a $V^R$ which diagonalizes
$A^R$ in this basis. In the CFL case, where the diagonalized 
$A^L$ is proportional to the identity, $a=b=c$, it is easy to show that 
one can choose $V^R$ such that $A^R = \pm A^L$. These two configurations
are related by a $U(1)_A$ rotation (see section 3). Hence, they are degenerate
in the high density limit where gluon exchange dominates. Instanton effects,
important at intermediate density, favor $A^R = A^L$.
Note that parity, if unbroken, requires 
$A^L = A^R$, and hence implies simultaneous diagonalizability.

In what follows we consider the potential vacua parametrized by a,b,c.
First, we use the Dyson-Schwinger (gap) equation to determine which of these 
configurations are energy extrema. Next, we compute the energies of the
extrema to determine the true groundstate. A similar analysis has been
carried out by Sch\"afer and Wilczek \cite{SW2} in the approximation where gluon
interactions are replaced by local four fermion
interactions. They concluded that the CFL vacuum had the lowest energy.
Here, we include the gluons in the analysis, introducing
long range color-magnetic fluctuations (controlled by Landau damping) 
and Meissner screening into the gap equation and vacuum energy calculations.

We find that the CFL vacua remains the lowest energy state,
at least at asymptotically high densities where the  calculation is reliable.
The Meissner effect is a small correction to the
vacuum energy at asymptotic densities. At lower densities where the 
gauge coupling is large the Meissner terms become more important and 
tend to disfavor CFL relative to the absence of a condensate. 
We do not know whether they are 
ever sufficient to remove the superconducting 
phase but they will lower the energy difference between the vacuum and 
unstable condensates with different color and flavor breaking patterns. 
Configurations which satisfy the gap equations but are not the global
minimum of energy are presumably saddlepoints, since they are continuously
connected to the CFL vacuum via color and flavor rotations.

\section{Gap Equation}

In this section we determine the subset of parameter space for which
our ans\"atz (\ref{ansatz1}) satisfies the gap equation. Because the
gap equation results from the extremization of the effective action, 
its solutions are energy extrema.
At asymptoticaly high densities (weak coupling) the diagrams (a)-(c) in 
figure 1 give the leading approximation to the effective action. 
Note that in these diagrams the quark propagators include the diquark
condensate (see (\ref{SI}) below), and the gluon propagators include
Landau damping, but {\it not} the Meissner effect. The latter arises
from the condensate-dependence of quark loops in diagrams (c) and (d). 
The resulting gap equation (figure 2, with condensate shown
explicitly at lowest order in $\Delta$) is given by
\beq
S^{-1}(q) - S^{-1}_0 (q) ~=~ i g^2 \int {d^4k \over (2 \pi)^4} 
  ~\Gamma^A_\mu  
~S(k)~ \Gamma^B_\nu  ~D^{\mu \nu}_{AB} (k-q)~,
\eeq
where 
\beq
\label{G}
\Gamma^A_\mu~=~ \left( \begin{array}{cc}
\gamma_\mu T^A  & 0 \\
0 & C (\gamma_\mu T^A)^T C^{-1}
\end{array} \right)~~~.
\eeq
$~D^{\mu \nu}_{AB}$ is the gluon propagator, include the
effects of Landau damping and Debye screening (we assume
Feynman gauge throughout):
\beq
\label{D}
D^{\mu\nu} ~=~{1\over q^2 + G} P_T^{\mu\nu} ~+~ 
{1\over q^2 + F} P_L^{\mu\nu}~~~,
\eeq
where $P_{T,L}$ are transverse and longitudinal projectors. The analytic
forms of F and G are given below in (\ref{FG}).
The small $\frac{q_0}{q}$ expansion of G leads to the Landau damped magnetic
gluon propagator
\beq
\label{LDD}
D_T^{\mu \nu} (q_0, q) 
~=~ { P_T^{\mu \nu} \over q^2 + i\frac{\pi}{2} m_D^2 \frac{|q_0|}{q} }~~~,
\eeq
while the expansion of F leads to the usual longitudinal propagator, with
Debye screening: $m_D^2 = N_f {g^2 \mu^2 \over 2 \pi^2}$.

We will restrict the color group structure in the gap equation
to the attractive anti-symmetric $\bar{3}$ channel:
\beq
T^A_{ab} T^A_{cd} ~ \rightarrow ~ {1 \over 3} \left(
\delta_{ac} \delta_{bd} - \delta_{ab} \delta_{cd} \right)~~~,
\eeq
which projects out the anti-symmetric part of $S(k)$ in color space in
the gap equation.
Here $S$ is the fermion propagator for the spinor $(\psi^i_a, \psi^{i C}_a)$ 
with $i$ a flavor
index and $a$ a color index.

\epsfysize=3 cm
\begin{figure}[htb]
\center{
\leavevmode
\epsfbox{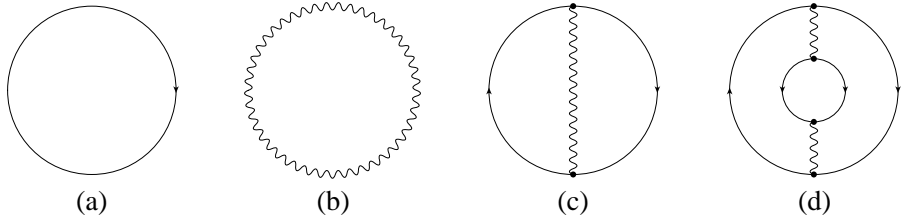}
\caption{Vacuum energy diagrams} \label{fig1}
}
\end{figure} 

\epsfysize=3 cm
\begin{figure}[htb]
\center{
\leavevmode
\epsfbox{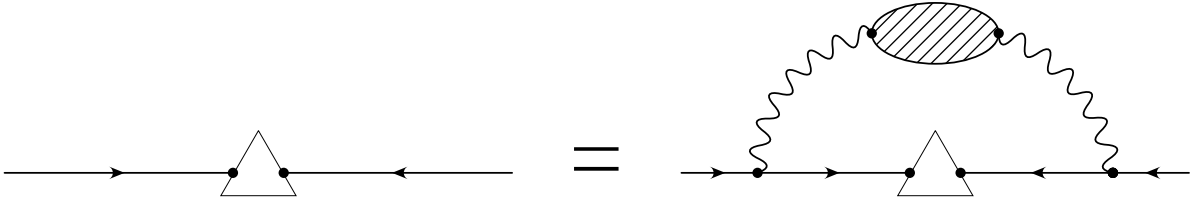}
\caption{Dyson-Schwinger equation} \label{fig2}
}
\end{figure}

For the three flavor case
$S$ can be written explicitly as an 
$18 \times 18$ matrix in color flavor space. 
The inverse propagator may be written
\beq
\label{SI}
S^{-1}(q) = \left( \begin{array}{cc}
q\!\!\!/ + \mu\!\!\!/  & \gamma_0 \Delta^\dagger \gamma_0 \\
\Delta  &  q\!\!\!/ - \mu\!\!\!/ 
\end{array} \right)
\eeq
where $\mu\!\!\!/ = \mu \gamma_0$.
$\Delta$ is a $9 \times 9$ matrix which for the ans\"atz (\ref{ansatz1}) 
takes the form
\beq \label{cond}
\Delta = 
\left( \begin{array}{ccccccccc}
   0 & 0 & 0 & 0 & c & 0 & 0 & 0 & b \\
   0 & 0 & 0 & -c & 0 & 0 & 0 & 0 & 0 \\
   0 & 0 & 0 & 0 & 0 & 0 & -b & 0 & 0 \\
   0 & -c & 0 & 0 & 0 & 0 & 0 & 0 & 0 \\
   c & 0 & 0 & 0 & 0 & 0 & 0 & 0 & a \\
   0 & 0 & 0 & 0 & 0 & 0 & 0 & -a & 0 \\
   0 & 0 & -b & 0 & 0 & 0 & 0 & 0 & 0 \\
   0 & 0 & 0 & 0 & 0 & -a & 0 & 0 & 0 \\
   b & 0 & 0 & 0 & a & 0 & 0 & 0 & 0 
 \end{array} \right)
\eeq
Because we are dealing with a diquark condensate
the non-trivial part of the gap equation involves the 
lower left $9 \times 9$ block. We will refer to this sub-block of the 
propagator $S$ as $S_{21}$. 

For a particular ans\"atz $\Delta$ to be a solution to the gap equation we 
require that the color antisymmetric part of 
$T^A~ S_{21}(k) ~T^A$ (corresponding to the $\bar{3}$ channel) 
be proportional in color-flavor space to 
$S^{-1}(q) - S^{-1}_0(q) = \Delta (q)$, 
which appears on the LHS of the gap equation. 
This requires some justification, as the matrices that appear
on the RHS of the gap equation appear inside the integral. 
If, for simplicity, we restrict ourselves to ans\"atze which
correspond to constant color-flavor matrices times a function of
momenta, then this condition is implied. In principle, there can
be more exotic solutions in which color and flavor orientations
rotate in momentum space, however it seems unlikely that such
solutions exist. We note that the
equality must hold for all values of the external momentum q, and that
the set of functions $D(k-q)$ are likely to form a complete basis for 
functions of k, since they are essentially smeared delta functions of 
(k-q). Thus by taking appropriate linear combinations of the gap equation
we can see that $T^A~ S_{21}(k) ~T^A$ must be proportional to $\Delta$
when integrated against any arbitary function of k. Hence the proportionality
must hold without the integral. 

The propagator may be found by inverting the
sparse matrix in (\ref{SI}) using Mathematica. 
Only three ans\"atze satisfy our condition:
$a=b=c$; $a=b, c=0$; $b=c=0$. We refer to these solutions as
(111) (color-flavor locking), (110) ($3 \rightarrow 0$ breaking) 
and (100) ($3 \rightarrow 2$ breaking) respectively.

For these ans\"atze the color antisymmetric part of 
$T^A S_{21}(k) T^A$ has the form of a constant
multiplying the matrix form (\ref{cond}) with $a,b,c$ set
to 0 or 1 as is appropriate for the ans\"atz. The constants are
(here $l^2 = (|\vec{k}| - \mu)^2$): 
\beqa
(111) ~&:&~ { 2 \Delta \over (k_0^2 - l^2 + \Delta^2) }~ { (k_0^2 - l^2 + 3 
\Delta^2) \over (k_0^2-l^2 + 4 \Delta^2) } \nonumber \\
(110) ~&:&~ { \Delta \over (k_0^2-l^2 + \Delta^2)}  + 
{ \Delta \over (k_0^2-l^2 + 2 \Delta^2)}  \nonumber \\
(100) ~&:&~ { 2 \Delta \over (k_0^2-l^2 + \Delta^2) }
\eeqa
The integral over $l$ can be performed by contour integration, yielding
the following gap kernels

\beqa
(111) ~&:&~ \frac{2}{3} {\Delta \over \sqrt{k_0^2 + \Delta^2} } +  \frac{1}{3}
{\Delta \over \sqrt{k_0^2 + 4 \Delta^2} } \nonumber \\
(110) ~&:&~ { \Delta \over 2 \sqrt{k_0^2 + \Delta^2} } +
{ \Delta \over 2 \sqrt{k_0^2 + 2 \Delta^2} } \nonumber \\
(100) ~&:&~ { \Delta \over \sqrt{k_0^2 + \Delta^2} }
\label{kernels}
\eeqa

Let us now simplify the gap equations.
We neglect $q_0$, 
as compared to $|\vec q|$, as well as anti-particle contributions 
(see, e.g., \cite {SWgap}, for details). We obtain 


\begin {eqnarray}
\label{finalgap}
\Delta(p_0) &=& \frac{g^2}{12\pi^2} \int dq_0\int d\cos\theta\,
 \left(\frac{\frac{3}{2}-\frac{1}{2}\cos\theta}
            {1-\cos\theta+(G+(p_0-q_0)^2)/(2\mu^2)}\right. \\
 & & \hspace{3cm}\left.    +\frac{\frac{1}{2}+\frac{1}{2}\cos\theta}
            {1-\cos\theta+(F+(p_0-q_0)^2)/(2\mu^2)} \right)
K(q_0), \nonumber
\end {eqnarray}
where
\begin {eqnarray}
F&=&2m^2\frac{q^2}{\vec{q}^{\,2}}\Bigg( 1-\frac{iq_0}{|\vec{q}|}
     Q_0\Bigg(\frac{iq_0}{|\vec{q}|}\Bigg)\Bigg),
\hspace{1cm} Q_0(x)=\frac{1}{2}\log\left(\frac{x+1}{x-1}\right), \nonumber \\
G&=&m^2\frac{iq_0}{|\vec{q}|}\Bigg[
    \Bigg(1-\Bigg(\frac{iq_0}{|\vec{q}|}\Bigg)^2\Bigg)
   Q_0\Bigg(\frac{iq_0}{|\vec{q}|}\Bigg) +
   \frac{iq_0}{|\vec{q}|} \Bigg],
\label{FG}
\end {eqnarray}
and $K(q_0)$ is one of the gap kernels from (\ref{kernels}).

These gap equations can be solved numerically. We first present
solutions neglecting the Meissner effect. 
The results for $\Delta$ vs $p_0$ are displayed in figures 3 and 4 
for the three ans\"atze\footnote{Our results differ somewhat in normalization
from those of \cite{SWgap}, although the shapes of the curves are 
in agreement.}. (The spatial momentum $\vec{p}$ is taken 
to lie on the Fermi surface.)
The curves lie very close to each other but as we
will see below give quite different contributions to the vacuum energy.
Note that the gap solutions we obtain have 
broad support, from the Fermi surface to $l, k_0 \sim \mu$. However,
this is likely a consequence of the approximations used in
(\ref{finalgap}), in which all momenta are assumed to lie close to the
Fermi surface.

\epsfysize=12 cm
\begin{figure}[htb]
\center{
\leavevmode
\epsfbox{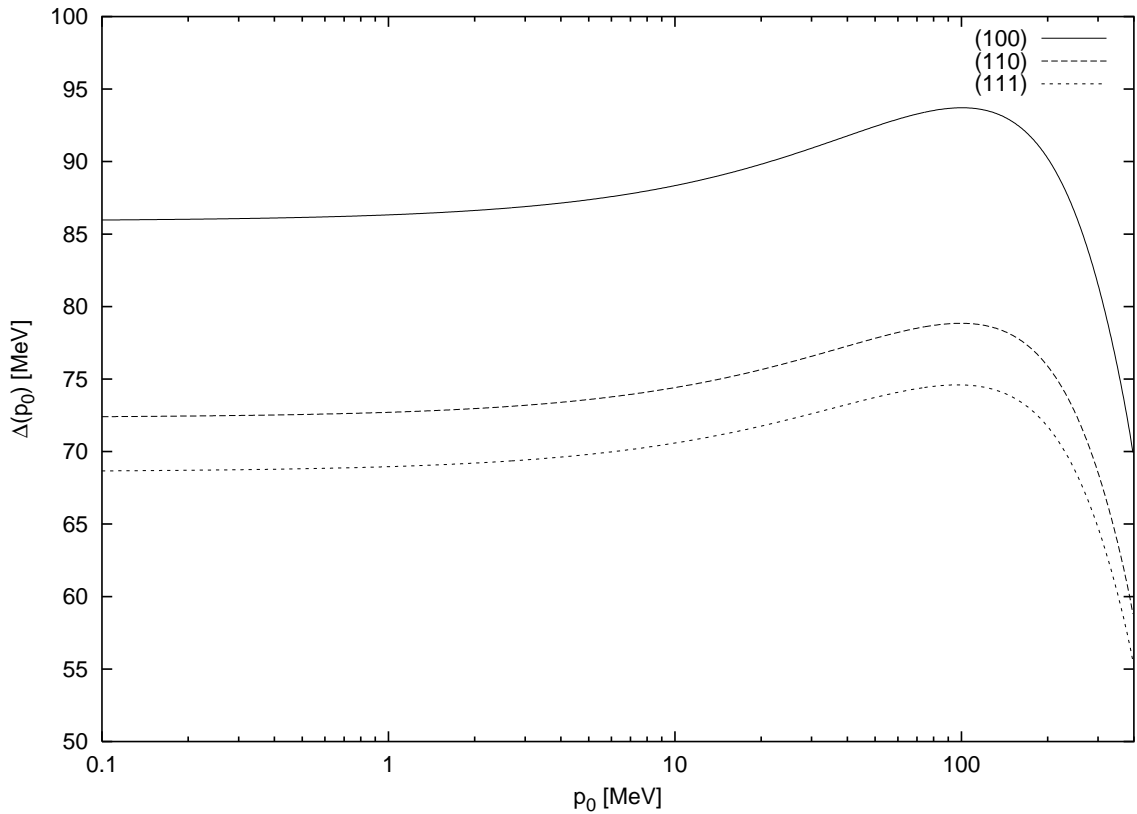}
\caption{Gap Solutions for $\mu = 400$ MeV } \label{fig3}
}
\end{figure}

\epsfysize=12 cm
\begin{figure}[htb]
\center{
\leavevmode
\epsfbox{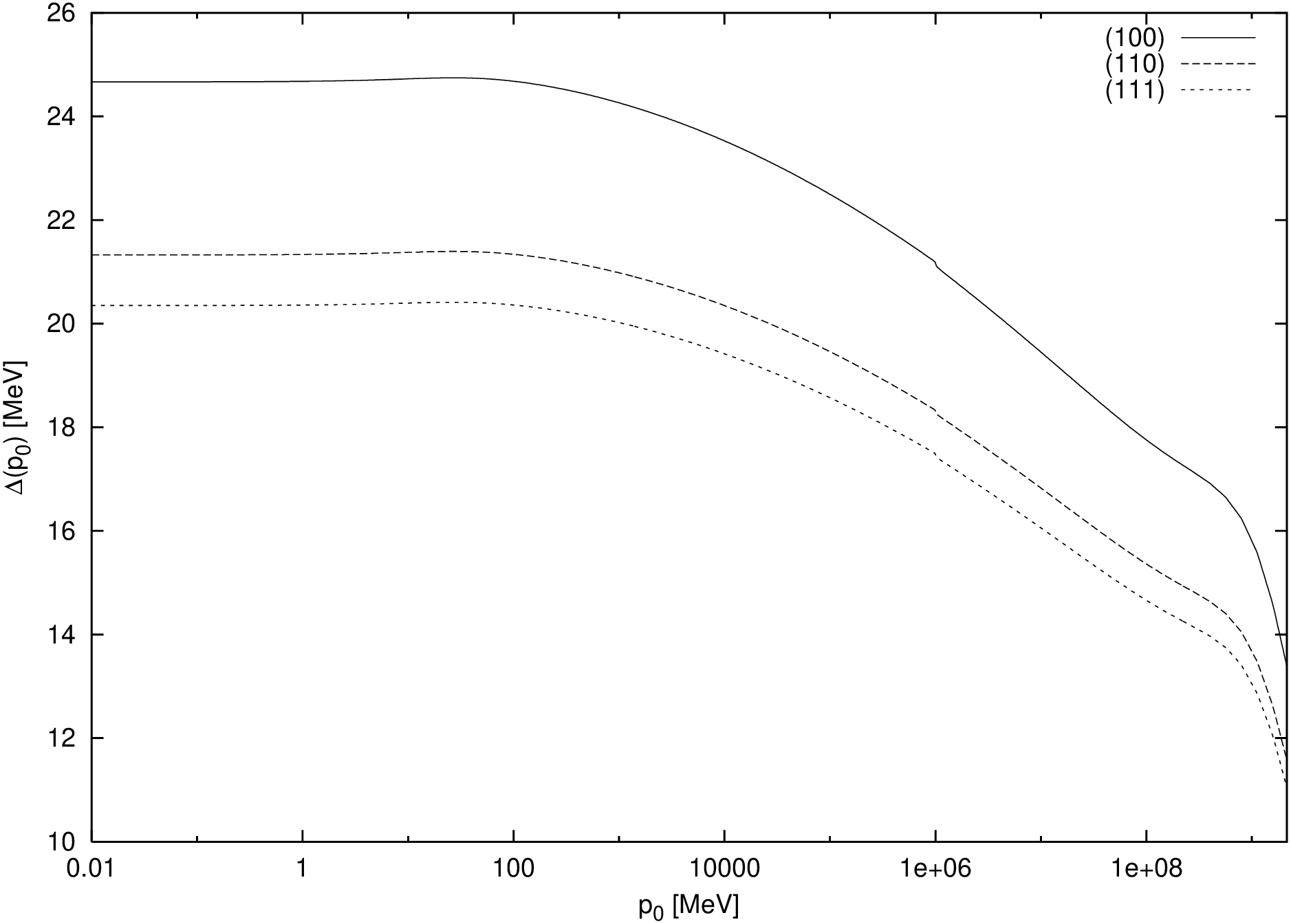}
\caption{Gap Solutions for $\mu = 10^{10}$ MeV} \label{fig4}
}
\end{figure}

A complete analysis must also include the Meissner effect, that is
the screening of the gluons induced by the formation of the gap.
The leading order contribution to the gluon vacuum polarization  $P(k_0,k)$
comes from the off-diagonal terms in the  fermion propagator-matrix
\beq
\label{P}
\delta P_{\mu\nu}(q_0,q) ~=~ g^2  \int {d^4k \over (2 \pi)^4} 
{\rm Tr} \Big[ \gamma_\mu \, S_{12}(k) \, C \gamma_\nu^T C^{-1} \, S_{21}(k-q) 
\Big] ~.
\eeq    
where we have dropped the group theory factors.
It is convenient to use the formalism of \cite{PRs} which exploits
simplifications due to decomposition of the fermion propagator into a sum of
projections onto different chirality and helicity channels.
After some tedious algebra one gets for (\ref{P})
\beq
\delta P_{ij}(q_0,q) ~=~ 2 g^2  
\int {d^4k \over (2 \pi)^4}\,  {|\Delta|^2 \over k_0^2 + \epsilon(k)^2}\,
{ g_{ij} + {1\over 2} (\hat{k}_i \widehat{k-q}_j + \hat{k}_j \widehat{k-q}_i)
\over (k_0 - q_0)^2 + \epsilon(k-q)^2}~,
\eeq
where $\hat{k} = \vec{k} / |\vec{k}|$ and 
$\epsilon(k)^2 = (|\vec{k}| - \mu )^2 + |\Delta |^2$. 
It is easier to compute the contribution to the magnetic gluon mass $G$
directly applying the transverse projector 
$P^T_{ij} = \Big(\delta_{ij} - \hat{q}_i \hat{q}_j\Big)$ to the
gluon vacuum polarization while using the HDL approximation (the 
momentum in the loop $k \sim \mu $ and is much bigger than the 
momentum transfer $q$) \cite{Bellac}:
\beq
\delta G(q_0, q)~=~ {1 \over 2} P^T_{ij} \delta P_{ij}(q_0,q) ~=~ 
g^2  \int {d^4k \over (2 \pi)^4}\,  
{|\Delta|^2 \over k_0^2 + \epsilon(k)^2}\,
{ 1 + \hat{k} \cdot \hat{q} \over (k_0 - q_0)^2 + \epsilon(k-q)^2}~~~.
\eeq
Further simplification comes after switching on a small temperature and
performing a summation over frequencies \cite{Bellac}. Note that, because
the system is already decomposed into particles and anti-particles about
the Fermi surface, one should apply the summation formulae as if 
$\mu =0$. Finally, one finds
\beq \label{self}
\delta G(q_0, q)~=~ 2\, {g^2 |\Delta|^2 \over (2 \pi)^3}
\int {d^3k \over \epsilon(k)\, \epsilon(k-q)} 
{ \epsilon(k) + \epsilon(k-q) \over q_0^2 + ( \epsilon(k) +
\epsilon(k-q))^2}
~~~.
\eeq
One can see, either by analytical approximation or numerical evaluation,
that $\delta G(q_0,q)$ is of order $m_D^2$ for $q_0 \sim q \sim \Delta$,
and falls off like $1/q_0$ or $1/q$ as either become large \cite{HongSD}. 
While this
is of the same order as Landau damping, numerical
evaluation shows that the Meissner contribution is somewhat smaller.

\begin{figure}[htb]
\center{
\leavevmode
\epsfbox{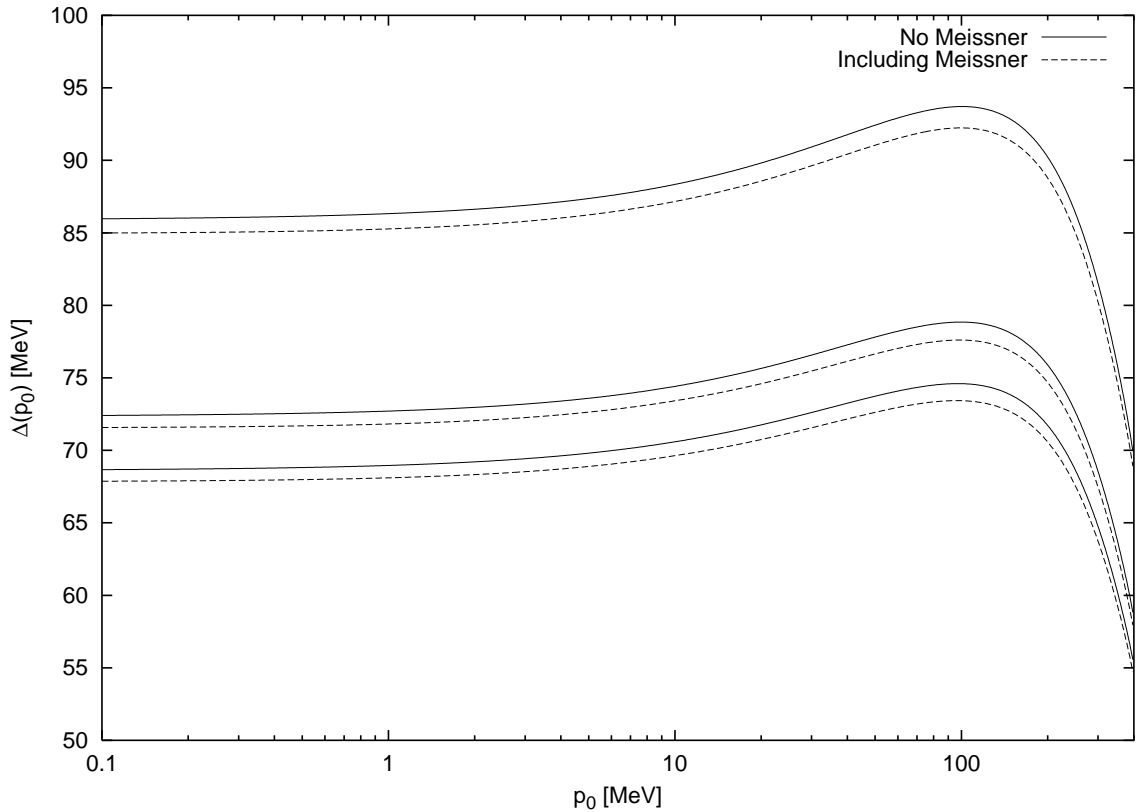}
\caption{Gap Solutions for $\mu = 400$ MeV } \label{fig5}
}
\end{figure}

\begin{figure}[htb]
\center{
\leavevmode
\epsfbox{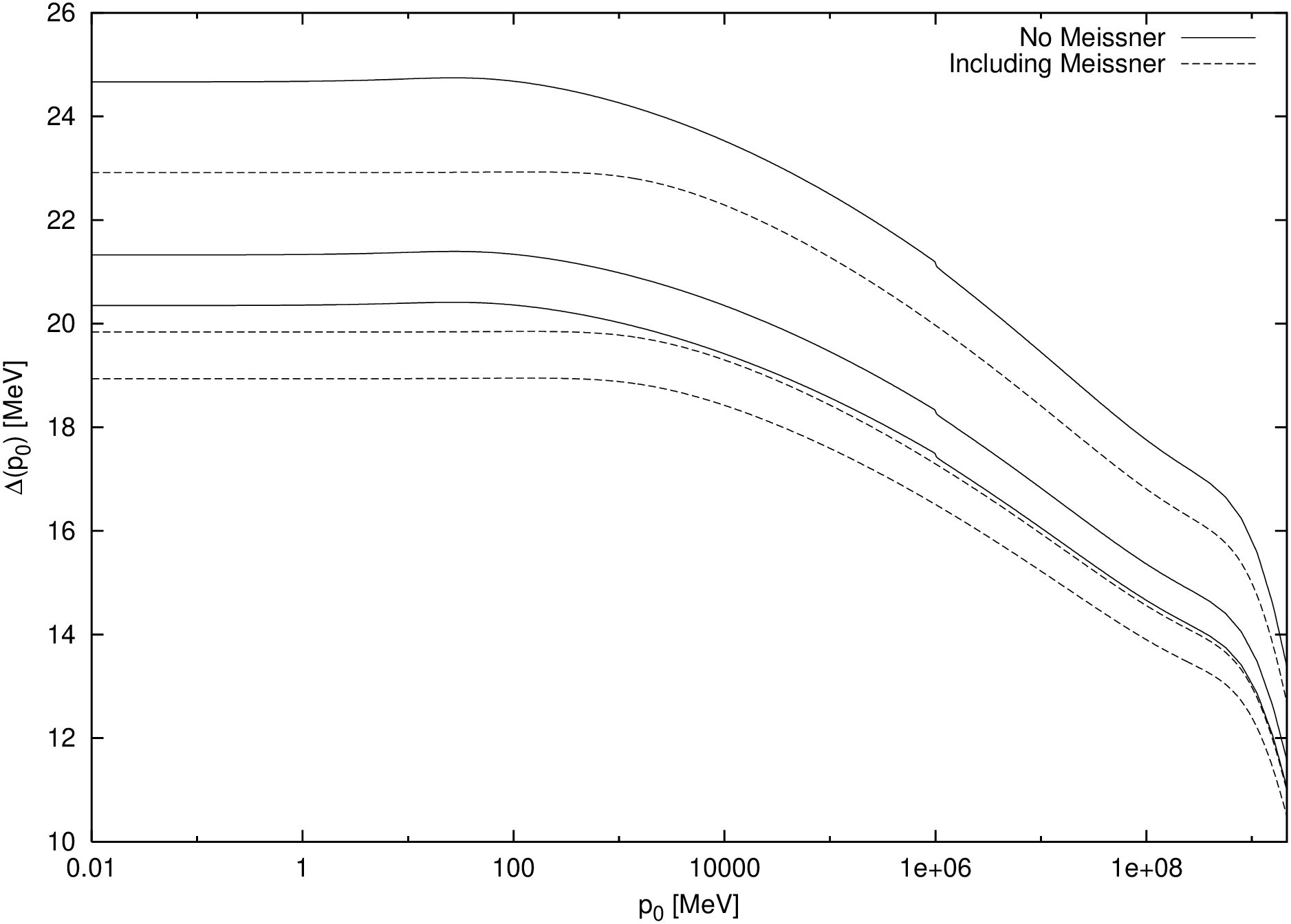}
\caption{Gap Solutions for $\mu = 10^{10}$ MeV} \label{fig6}
}
\end{figure}

As we are only interested in the size of the contribution of the Meissner
effect, we use the following approximation, which is an overestimate
of the effect: 
\beq
\delta G(q_0,q) \simeq m_D^2  {\Delta_0 \over  \sqrt{ q^2 + q_0^2 +
\Delta_0^2 } } ~~~ ,
\eeq
where $\Delta_0$ is the maximum value of the function $\Delta(k_0,k)$.
The gap equations were numerically solved for all three gap kernels, and 
the results are shown in figures \ref{fig5} and \ref {fig6}.  The effect
is to decrease the size
of the condensate but it is a small perturbation on the solutions obtained 
previously.

\epsfysize=12 cm

\epsfysize=12 cm

\section{Vacuum Energies}

To determine which of the above gaps is the true minimum energy state we 
must calculate the vacuum energy, which receives contributions from 
vacuum to vacuum loops of both quarks and gluons (figure 1). 
We use the CJT effective potential, which is 
a function of condensates \cite{CJT}:
\beqa
\label{cjt}
V(S,D)~&=&~ -i \int {d^4p \over (2 \pi)^4} \Big[ {\rm tr} \ln S(p)/S_0(p)
~+~ {\rm tr} (1 - S(p)/S_0(p))
~-~{i\over 2}  {\rm tr} \ln D(p)/D_0(p) \nonumber \\
~&-&~ {i\over 2} {\rm tr} (1 - D(p)/D_0(p))\Big]
~+~ {i\over 2} \int \int {d^4p \over (2 \pi)^4}{d^4k \over (2 \pi)^4}
\Big[ {\rm tr} \Gamma S(p) \Gamma S(p+k) D(k) \Big] \nonumber \\
~&+&~ \cdots
\eeqa
where for convenience we suppress appropriate color, flavor and Dirac 
indices. $S_0$ and $S$ correspond to bare and full fermion propagators, 
$D_0$ and $D$ to bare and full gluon propagators and $\Gamma$ to full
vertices.
The ellipsis denote gluon self-interaction loops and 
terms which are higher order
in g. In our approximation, which is essentially Hartree-Fock 
(lowest order in coupling), the $\Gamma$'s become bare vertices. 

Extremizing with respect to appropriate propagators and 
vertices one obtains a set of gap equations. The fermion gap equation
is the one we studied in the previous section, while the gluon gap
equation produces Landau damping. We wish to 
compare values of $V(S,D)$ corresponding to our three solutions to 
determine which one is the true vacuum\footnote{The difference in
energies $V$ will be gauge invariant, whereas actual values are not.}. 
It is easy to show that the value of the effective potential 
evaluated on the gap solution $(S^*, D^*)$ 
in the Hartree-Fock approximation is given by:
\beq
\label{cjtg}
V(S^*,D^*)~=~ -i \int {d^4p \over (2 \pi)^4} {\rm tr ln} S(p)/S_0(p)~.
\eeq
Diagramatically, this is equivalent to the graph of figure 1(a) when
evaluated on the gap solution.

The fermion
loops are most easily calculated by going to a basis where $S_0 S^{-1}$ 
is diagonal in color-flavor space. Note that the gap matrix $\Delta$
has non-trivial Dirac structure that must be accounted for \cite{PRs}:
\beq
\Delta = \Delta_1 \gamma_5 P_+ ~+~ \Delta_2 \gamma_5 P_-~~~,
\eeq
where $P_{\pm}$ are particle and anti-particle projectors. Our analysis
has been restricted to the particle gap function $\Delta_1$. The anti-particle
gap function $\Delta_2$ has its support near $k_0 \sim 2 \mu$, 
and its contribution to the vacuum energy is suppressed.
There are 18 eigenvalues, which occur in 9 pairs. The 
product of each pair is of the form 
\beq
 - \left( 1  ~+~ a { \Delta^2 ( k_o, k ) \over 
k_0^2 ~+~ (|\vec{k}| - \mu )^2 }  \right) 
~~~,
\eeq
where a is an integer. For our three cases we obtain the following
sets of eigenvalues:
\beqa
(111) ~~ &\rightarrow& ~~  8 \times \{ a=1 \} ~,~ 
1 \times \{a = 4 \} \nonumber \\ 
(110) ~~ &\rightarrow& ~~  4 \times \{ a=1 \} ~,~ 2 \times  \{ a = 2  \} 
\nonumber \\
(100) ~~ &\rightarrow& ~~  4 \times \{ a=1 \}
\eeqa

The binding energy is of order
\begin{eqnarray}
\label{Equark}
E_q & \sim &  - \int d^3k  dk_0 ~
\ln \left[ 1  + a { \Delta^2 (k_0,k) \over k_0^2 + (k - \mu)^2 } \right] \\
& \sim & - a ~\mu^2 \Delta_0^2 ~~, 
\end{eqnarray}
where $\Delta_0$ is the maximum value of the gap function $\Delta (k_0,k)$,
which has rather broad support in both energy and momentum space away from
the Fermi surface, extending to $k_0, k \sim \mu$. A more precise answer
than (\ref{Equark}) requires numerical evaluation, but it is clear that the
result scales with a and has only a weak (logarithmic) dependence on the
variations in the shape of $\Delta (k_0,k)$. Substituting our numerical
results for the gaps in the three cases, it is easy to establish that
\beq
E(111) ~<~ E(110) ~<~ E(100)~~~.
\eeq

Gluon loops corresponding to figure 1(b) yield a smaller contribution
to the vacuum energy. To compute this energy we must use the
gluon propagator suitably modified by the Meissner effect, which as
we described above leads to the vacuum polarization $P(k_0,k)$. We
obtain
\begin{eqnarray}
E_g & = & {3 \over 64 \pi^3} \int dk_0 dk ~k^2 ~\ln 
\left[  1 + { P(k_0,k) \over  k_0^2 + k^2 } \right]~~~.
\end{eqnarray}
To estimate the result of this integral it is necessary to use
the properties of $P(k_0,k)$. Recall that $P(k_0,k)$ falls off
like $m_D^2 \Delta_0 /k$ at $k >> \Delta_0$, and similarly at large
$k_0$ \cite{HongSD}. The dominant region of integration is therefore 
$k_0 \sim \Delta_0$ and $k_* << k << \mu$, where 
$k_* = m_D^{2/3} \Delta_0^{1/3}$ is the momentum scale familiar from
Landau damping. From this region of integration we obtain
\beq
E_g ~\sim~ m_D^2 \Delta_0^2 \ln( \mu / k_* ) ~\sim~ g \mu^2 \Delta_0^2~~~,
\eeq 
since $\ln k_* \sim 1/g$. The result is parametrically smaller than the 
quark contribution. Note that this contribution to the energy is
positive and hence prefers the least possible breaking of the color
gauge symmetry. If this term were the dominant one then it would
disfavor the  formation of a condensate and the CFL vacuum
would be the highest energy state! At asymptotic densities it is
not the dominant term and the analysis from the quark loops
stands. At lower densities, where the coupling is large, these 
contributions to the energy become more important but we lose control 
of the calculation. Their effect is to lower the energy gap between the 
CFL vacuum and disoriented states with different color and flavor symmetry 
breaking patterns.

The contribution we just computed to $E_g$ still
cannot differentiate between relative color rotations between the LL and RR 
diquark condensates. This is because the energy $E_g$ depends on the
sums of the squares of the gauge boson masses induced.
In order to be sensitive to LR coupling effects, it
is necessary that both LL and RR contributions to $P(k_0,k)$ appear 
simultaneously in the vacuum energy contribution. 
The first such graph is that of figure 1(d),
and it is of the form
\beq
E_g^{LR} \sim \int dk_0 dk~ k^2~ \ln \left[ 1 + 
\left({ P(k_0,k) \over  k_0^2 + k^2 }\right)^2 \right]~~~.
\eeq 
This integral is dominated by the region $k_0 \sim \Delta_0, k \sim k_*$,
leading to the result
\beq
E_g^{LR} ~\sim~ m_D^2 \Delta_0^2 ~\sim~ g^2 \mu^2 \Delta_0^2~~~. 
\eeq
This effect is further suppressed by a power of the coupling constant.
We see that in the weak coupling limit the vacuum energy required to 
disorient the LL and RR condensates in color space is rather small. This
suggests that even at asymptotic densities 
in the two flavor case it might be possible to 
disorient the diquark condensates from their lowest energy configuration.
We have yet to determine what this lowest energy configuration is, and hence
whether parity is violated in the two flavor case. In principle, one should
minimize $E_g^{LR}$ as a function of the relative LL and RR color orientation.
Instead, we will give a simple argument that the condensates prefer to align.
We noted in the last section that including the Meissner screening in the
gluon propagator leads to a decrease in the gap size. This is a small
effect at weak coupling, and was negligible compared to the color-flavor 
structure of the quark propagator. However, in determining LL-RR alignment 
it is the main effect. 
In the two flavor case none of the gluons responsible for the 
attractive interaction are Meissner screened, as long as the LL and RR 
condensates align. That is the quarks which condense are those that 
transform under the unbroken SU(2) subgroup of $SU(3)_c$.
However, any misalignment leads to the LL condensate
screening the RR channel and vice versa, decreasing the condensates and
thereby increasing the energy. Hence in the two flavor case the condensates 
prefer to align and parity is preserved. In the three flavor case CFL 
gives all of the gauge bosons a mass and this effect is absent. 

In both the two and three flavor cases there remains the possibility of 
parity violation through a phase associated with the $U(1)_A$ symmetry
\cite{ARW1,EHS,PR}. Only instanton effects (highly suppressed at asymptotic 
densities) can distinguish these vacua. At lower densities 
instanton effects are expected to strongly break the $U(1)_A$ symmetry,
since the $\eta'$ mass is dominated by these effects at zero density.

\section{Conclusions and Phenomenology}

In this paper we analyzed the possible ground states of QCD at
asymptotic densities. We verified that in the two flavor case, the
symmetry breaking pattern is $\rm SU(3)_c \rightarrow SU(2)_c$,
while in the three flavor case, color flavor locking has the lowest
vacuum energy. In neither case is parity spontaneously violated 
until the density is strictly infinite \cite{EHS,PR}. 

Our analysis of the energy surface governing color superconductivity
suggests possible experimental signatures in heavy ion collisions. 
In particular the existence of relatively flat directions along which
color and flavor symmetry breaking patterns change 
raises the possibility of domains of disoriented condensates, each with
distinct hadronization properties. In the two flavor
case the LL and RR condensates, each of which break 
$SU(3)_c \rightarrow SU(2)_c$,
are only aligned by a subleading term in the vacuum energy calculation. In a 
heavy ion collision we might expect the condensates to be 
misaligned by an arbitrary $SU(3)_c$ transformation, leading to violation of
parity and complete breaking of the color group. In the three flavor
case we might expect much the same.
Here the gauge loop contributions to
the vacuum energy from gluon loops will tend to reduce
the energy difference between the CFL and, for example, the 
(1,0,0) condensates as
discussed above. The strange quark mass also tends to reduce the energy
gap between these two condensates as discussed in \cite{ARW2}.
For some (uncalculable) value of $m_s$  of order $\Lambda_{QCD}$
we expect a phase transition between these two condensates as the  
number of light flavors changes from 3 to 2. Thus for realistic
values of $m_s$, and densities and temperatures achievable in heavy
ion collisions, we might expect disoriented condensates to form with
a range of possible color symmetry breaking patterns appearing 
on a collision by collision basis.

To see how such variation in color symmetry breaking might be seen 
in an experiment we consider the extreme case where the $SU(2)_c$ subgroup
is left unbroken (this is the true vacuum of the two flavor theory). Consider
a region which in the wake of a heavy ion collision volume is sufficiently 
cool and dense to allow the formation 
of a diquark condensate, with gauge symmetry broken to $SU(2)_c$. 
The region presumably expands and cool in the usual fashion. However,
one color of quark (e.g. red) does not participate in the condensation and
its propagator is unaffected by $\Delta$. 
It is also more weakly interacting since its color corresponds
to precisely the broken part of the gauge group (gluons which couple
to red quarks are screened by the Meissner effect). 
The remaining two colors of quarks participate in Cooper pairing 
and interact strongly with the plasma, so they do not disperse as quickly. 
The red quarks will therefore tend to flow to the surface of the fireball,
providing a mechanism for macroscopic transport of color charge.
Note that the condensate is stable under this
charge separation since it is the condensate 
favored by an $SU(2)_c$ theory with 
two flavors. Furthermore, with $SU(3)_c$ broken to $SU(2)_c$ there is no
restoring force which prevents this charge polarization.
On leaving the superconducting volume
red quarks will suddenly be required to 
hadronize because their color charge can now support long range fields and
they become aware of the large value of the other 
two color charges in the center. We expect that this color polarized
fireball will hadronize very differently 
than a quark gluon plasma which is locally color neutral.
Naively one expects quark anti-quark production on the boundary 
of the color charge separation in order to enforce charge neutrality.
The separated red charge would then emerge as 
energetic hadrons, leaving a cooler central region behind.

The scenario described above is the extreme case of a 
fully unbroken $SU(2)_c$ subgroup.
On the other hand a CFL state treats all colors equally and there will
be no charge polarization. On an event by event basis we expect variation
between these two extremes. The most likely signal of such events is a 
departure from the standard thermal distribution so far observed in 
heavy ion collisions \cite{thermal}, both on an event by event basis and
averaged over many events.

\bigskip \noindent
{\bf Acknowledgements:} SH would like to thank D.K. Hong, V. Miransky, 
M. Nowak and M. Rho for useful discussions. Part of this work was
performed at the Korean Institute for Advanced Study (KIAS). SH is
supported under DOE contract DE-FG06-85ER40224.
NJE acknowledges the support of a PPARC Advanced Fellowship.
MS is supported under DOE contract DE-FG02-91ER40676. JH
is supported in part by the Natural Sciences and Engineering Research 
Council of Canada and the Fonds pour la Formation de
Chercheurs et l'Aide \`a la Recherche of Qu\'ebec.

\end{document}